\documentclass{article}

    \makeatletter
\def\@fnsymbol#1{\ensuremath{\ifcase#1\or \dagger\or \ast\or
   \mathsection\or \mathparagraph\or \|\or **\or \ddagger\ddagger
   \or \ast\ast \else \fi}}
    \makeatother
            
\usepackage{PRIMEarxiv}
\usepackage[labelfont=bf]{caption}
\usepackage{amsmath}
\usepackage[utf8]{inputenc} 
\usepackage[T1]{fontenc}    
\usepackage{hyperref}       
\usepackage{url}            

\usepackage{booktabs}       
\usepackage{amsfonts}       
\usepackage{nicefrac}       
\usepackage{microtype}      
\usepackage{lipsum}
\usepackage{fancyhdr}       
\usepackage{graphicx}       

\vspace{-10mm}

\title{SELFormer: Molecular Representation Learning via SELFIES Language Models}

  \author{
  Atakan Yüksel\thanks{Equally contributing co-first authors.}	  \\
  Biological Data Science Lab \\ 
  Dept. of Computer Engineering \\
  Hacettepe University 
   \And
  Erva Ulusoy$^\dagger$ \\
  Biological Data Science Lab / \\ 
  Bioinformatics Department \\
  Hacettepe University 
   \And
   Atabey Ünlü$^\dagger$ \\
   Biological Data Science Lab / \\
   Bioinformatics Department \\
   Hacettepe University 
   \And
    Tunca Doğan\thanks{Corresponding author email address: tuncadogan@gmail.com}\\
   Biological Data Science Lab  \\ 
   Dept. of Computer Engineering\\
   Hacettepe University 
}

\begin{document}

\maketitle

\begin{abstract}
Automated computational analysis of the vast chemical space is critical for numerous fields of research such as drug discovery and material science. Representation learning techniques have recently been employed with the primary objective of generating compact and informative numerical expressions of complex data, for efficient usage in subsequent prediction tasks. One approach to efficiently learn molecular representations is processing string-based notations of chemicals via natural language processing (NLP) algorithms. Majority of the methods proposed so far utilize SMILES notations for this purpose, which is the most extensively used string-based encoding for molecules. However, SMILES is associated with numerous problems related to validity and robustness, which may prevent the model from effectively uncovering the knowledge hidden in the data. In this study, we propose SELFormer, a transformer architecture-based chemical language model that utilizes a 100\% valid, compact and expressive notation, SELFIES, as input, in order to learn flexible and high-quality molecular representations. SELFormer is pre-trained on two million drug-like compounds and fine-tuned for diverse molecular property prediction tasks. Our performance evaluation has revealed that, SELFormer outperforms all competing methods, including graph learning-based approaches and SMILES-based chemical language models, on predicting aqueous solubility of molecules and adverse drug reactions, while producing comparable results for the remaining tasks. We also visualized molecular representations learned by SELFormer via dimensionality reduction, which indicated that even the pre-trained model can discriminate molecules with differing structural properties. We shared SELFormer as a programmatic tool, together with its datasets and pre-trained models. Overall, our research demonstrates the benefit of using the SELFIES notations in the context of chemical language modeling and opens up new possibilities for the design and discovery of novel drug candidates with desired features.

\end{abstract}

\newpage

\vspace{-4pt}

\section{Introduction}

\vspace{-8pt}

Traditional methods and techniques of drug discovery are expensive, resource-intensive, and time-consuming, making it impractical to study numerous molecules and identify those that could be developed into new drugs (Vamathevan et al., 2019). Incorporating novel computational methods into research and development pipelines is crucial, particularly in light of recent events that have highlighted the need for prompt action against sudden disease outbreaks (Basu et al., 2021). Molecules must be encoded and stored using alphanumeric characters in order to be processed computationally. SMILES (Simplified Molecular-Input Line-Entry System) is a notation for expressing the chemical structure of a molecule using a standard string of linear characters (Wang et al., 2019). Although SMILES is widely used and has proven useful in computational chemistry, it is insufficient to fully represent the complexity and properties of molecules. Not every SMILES string corresponds to a valid molecular graph, thereby reducing the representational space (Krenn et al., 2020). In order to create a better notation for molecules, IUPAC designed InchI, which has a canonical representation that assigns a unique string to each molecule. The InchI system encodes additional information about molecular graphs, including mobile and immobile hydrogens. However, InchI strings are difficult to comprehend, and their complex syntax makes their use in generative modeling challenging (Krenn et al., 2022). SELFIES is a novel method for representing molecular graphs as character strings that permits the unique identification of any given molecule (Krenn et al., 2020). SELFIES can be used for multiple purposes, including the construction of molecular fingerprints, the calculation of molecular similarity, and the detection of chemical reactions. SELFIES is human-readable and 100\% robust, meaning that every SELFIES string points to a valid molecule, making it more suitable for molecular representation learning than SMILES (Krenn et al., 2020). SELFIES have been utilized for drug discovery-related tasks in multiple studies (Nigam et al., 2021; Frey et al., 2022). \\

\vspace{-5pt}

Entities/objects (e.g., molecules) must be represented numerically to be utilized in the artificial learning-based analysis. These representations can be manually crafted (i.e., created empirically using pre-defined rules or descriptions) or automatically extracted in the context of machine learning. Representation learning is concerned with automatically discovering meaningful data representations (Bengio et al., 2013). These representations are intended to capture the underlying structure and patterns in the data. Some of the primary advantages of representation learning are that; (i) it can help reduce the dimensionality of the data, making it easier to manipulate and interpret (Kopf \& Claassen, 2021); and (ii) it generates generalized vectors that are potentially reusable in a variety of tasks, such as classification, clustering, and generation (Unsal et al., 2022). In recent years, numerous examples of representation learning strategies have been published (Chuang et al., 2020). Self-supervised learning is one example in which representations are learned from large amounts of unlabeled data using pretext tasks such as predicting the next token in an input sequence. These learned representations can then be fine-tuned for specific tasks, such as object recognition or sentiment analysis, for which smaller amounts of labeled data is sufficient (Wang et al., 2022a; AlBadani et al., 2022). Overall, representation learning has emerged as an important subfield of artificial learning, as it improves performance and applicability across a broad range of tasks (Ericsson et al., 2022). \\

\vspace{-5pt}

In recent years, transformers have become a widely used architecture for learning self-supervised text representations in the field of natural language processing (NLP) (Vaswani et al., 2017). Transformers achieved state-of-the-art results on many NLP tasks, including machine translation, text summarization, and question answering. The transformer is based on the concept of self-attention, which allows the model to consider the intrinsic relationships between different words or tokens in a sentence or sequence (Kalyan et al., 2021). This allows the transformer model to capture long-range dependencies in the data (Tay et al., 2022). The resulting encodings/embeddings  are used as the learned representation of the given sequence (Vaswani et al., 2017). The transformer-based language modeling approach has a great deal of potential for automatically learning the properties of molecules in large chemical libraries, which can then be used in various steps of drug discovery and development. However, the majority of the proposed chemical language models have a significant shortcoming in that they use SMILES notations to represent molecules. Problems with SMILES-based representations can be listed as; (1) there are a number of ways to write the same molecule in a non-canonical way, which decreases the uniqueness of molecular strings; (2) a valid SMILES string might have invalid chemical properties such as exceeding the natural valency of an atom (Wigh et al., 2022); (3) SMILES cannot fully capture spatial information; and (4) SMILES alone may not be sufficient to capture molecular characteristics since it lacks syntactic and semantic robustness (Li et al., 2022b; Krenn et al., 2022). These issues suggest a new research direction that may produce chemical language models with a greater degree of generalizability. Therefore, we pose the following question: "Is it possible to obtain a better molecular representation using a chemical language (i.e., a string notation) that is more expressive and suitable for machine learning-based applications than SMILES?" \\ 

In this study, we propose a chemical language model that employs SELFIES representations of molecules in order to obtain their concise, flexible, and meaningful representations for use in a variety of downstream molecular tasks in drug discovery and development. We pre-trained our model, SELFormer, on ~2 million drug-like small molecules in the ChEMBL database, within the framework of self-supervised learning, and evaluated the quality of the output representations in various molecular property prediction-related benchmarks. We provide the method as a programmatic tool, together with trained and ready-to-use models and pre-calculated representation vectors for all ChEMBL molecules. We expect that SELFormer, along with the obtained results, will be useful to computational chemistry researchers in order to facilitate the development of effective solutions for the next generation of drug development.

\section{Related Work}

BERT (Devlin et al., 2019) is the first major deep learning-based language model that utilizes the transformer architecture. BERT uses two pre-training strategies: masked language modeling (MLM) and next sentence prediction (NSP). In the MLM task, BERT masks 15\% of the words in each input sequence by replacing them with a [MASK] token. The model then attempts to predict the original values of these masked words based on their semantic relationships with the non-masked words, which involves adding a classification layer on top of the encoder output and calculating the probability of each word in the vocabulary using the softmax function. The cross-entropy loss function only takes into account the prediction of the masked values, ignoring the prediction of the non-masked values. NSP is a binary classification task in which BERT receives sentence pairs as input and learns to predict whether the second sentence is the one that follows the first sentence in the original input. BERT achieved state-of-the-art results on tasks such as question-answering and language interference. MolBert (Fabian et al., 2020) is a chemical language model designed to learn high-quality and flexible molecular representation using the BERT model and SMILES notations of molecules as the input language. MolBert is pre-trained with the Guacamole benchmark (Brown et al., 2019) dataset which is curated from ChEMBL (Gaulton et al., 2017) consisting of ~1.6 million molecules. The model was fine-tuned and evaluated on MoleculeNet benchmarking (Wu et al., 2018) tasks, i.e., BACE, BBBP, and HIV datasets for classification tasks, and ESOL, FreeSolv, and Lipo datasets for regression. MolBert was tested on three self-supervised tasks: (i) masked language modeling, (ii) predicting the equivalence of two given SMILES inputs where the second SMILES can either be a synonymous permutation of the first one or a SMILES selected from the training data randomly, and (iii) predicting physicochemical properties of the given molecules. The authors found that MolBert achieved high performance on virtual screening and QSAR benchmarks. \\

RoBERTa (Liu et al., 2019) is a BERT-based model that introduces slight improvements over BERT. RoBERTa utilizes dynamic token masking instead of BERT’s static token masking, meaning that different tokens are masked in each epoch during training. RoBERTa also removed the next sentence prediction (NSP) task, arguing that it may have a negative effect on the performance by preventing models from learning long-range dependencies. RoBERTa outperformed BERT on the majority of the NLP tasks that were evaluated. ChemBERTa (Chithrananda et al., 2020) is a model based on RoBERTa and SMILES notations of molecules. The authors of the study investigated how the molecular property prediction performance of transformer-based architectures varies with the size of the pre-training dataset, using masked token prediction as the pre-training objective. The model was pre-trained on 77 million unique molecules curated from PubChem (Kim et al., 2023) and fine tuned on MoleculeNet tasks (Wu et al., 2018). The authors found that ChemBERTa did not outperform the baseline models on most of the downstream tasks, although it did show state-of-the-art performance on toxicity prediction (Tox21). They observed that masked language modeling had a positive impact on downstream task performance. The authors improved the pre-training procedure and achieved higher scores on the same benchmarking tasks in the new version of their model, ChemBERTa-2 (Ahmad et al., 2022). \\

Translating the InChI (Handsel et al., 2021)  is a method for predicting the IUPAC name of a chemical from its standard International Chemical Identifier (InChI) using a sequence-to-sequence machine learning model. In lieu of tokenizing the input and output into words or sub-words, this model analyzes the InChI and predicts the IUPAC name character by character. The model achieved a 91\% accuracy; however, its results were less precise for inorganic compounds compared to organic molecules. \\ 

ChemFormer (Irwin et al., 2022) is a transformer-based model based on the BART language model. While the BERT model uses masked language modeling for pre-training, BART utilizes denoising of the corrupted sequences by random tokens (Lewis et al., 2019). The authors pre-trained their models on 100 million SMILES strings curated from the ZINC database. ChemFormer was fine-tuned on downstream tasks such as sequence-to-sequence, regression, synthesis, and retrosynthesis. While sequence-to-sequence tasks are analogous to pre-training; lipophilicity, ESOL, and FreeSolv were used for regression experiments. The synthesis-related task takes two reactants as input and predicts the outcome of the reaction, while the retrosynthesis task was designed as its opposite. The authors reported competitive results against other SMILES and graph-based models. \\

MolFormer (Ross et al., 2022) is a chemical language model that creates molecular representations using SMILES strings of a large molecule dataset curated from the ZINC and PubChem databases. MolFormer was trained with 1.1 billion molecules and then fine-tuned for several classification (e.g., BBBP, Tox21, ClinTox, HIV, BACE, and SIDER) and regression (e.g., QM8, QM9, ESOL, FreeSolv, and lipophilicity) tasks. The model improves on linear attention units (Su et al., 2021) to speed up the training and lower the memory requirements. In the linear attention unit, MolFormer utilizes rotary positional embeddings instead of absolute positional embeddings, which was previously proposed in the RoFormer (Su et al., 2021) study. In return, MolFormer attained improved convergence of the training loss and enhanced stability. MolFormer outperformed all compared models on regression tasks and on half of the classification tasks. The authors concluded that large chemical language models perform better than geometric (i.e., graph-based) deep learning models and that, even though no topological information has been fed to the model, it could still distill such knowledge. MolFormer is an important step in creating a large language model; however, MolFormer’s large size may limit its accessibility, especially considering the average user/lab. A similar transformer-based method trained via masked language modelling (MLM) on 1.1 billion molecules, X-MOL (Xue et al., 2021), also demonstrated the potential of using large-scale unlabelled data for the in silico drug development. \\ 

In addition to chemical language models, there are methods for learning the representation of molecules that employ molecular graphs. In this approach, atom and bond information are generally encoded as nodes and edges  of a molecular graph, respectively. In studies such as SchNet (Schütt et al., 2017) and MGCN (Lu et al., 2019), graph representations were utilized to comprehend quantum interactions of small molecules, marking the beginning of graph learning applications to small molecules. \\ 

D-MPNN (also known as ChemProp) is a graph-based deep learning model designed to produce molecular embeddings as well as evaluate their properties based on supervised downstream tasks (Yang et al., 2019). In this study, the authors developed directed message-passing neural networks (D-MPNN) by enhancing the MPNN algorithm (Gilmer et al., 2017) in order to generate embeddings based on the graph representations of input molecules. The objective of the message-passing algorithm is to determine the properties of the entire graph by aggregating data from connected nodes. ChemProp reported model performance in terms of quantum mechanics, physical chemistry, biophysics, and physiology datasets in the context of downstream classification and regression tasks. On each benchmark, ChemProp outperformed (or competed with) its predecessors. The D-MPNN algorithm is a crucial step in the construction of meaningful molecular graph embeddings. There are models that utilize the modified versions of the MPNN algorithm to achieve similar goals, such as DimeNet (Gasteiger et al., 2020) and GeomGCL (Li et al., 2022a).  While DimeNet represents molecules based on atomic distances, GeomGCL utilizes both 2-D and 3-D information. GeomGCL further improved its model with the addition of contrastive learning, which was reported to be beneficial in the case of  insufficient labeled data. In their study, Hu et al. (2020) developed a new strategy for pre-training GNNs at the level of both individual nodes and entire graphs simultaneously, and addressed the difficulties of graph pre-training when task-specific labels are scarce. Other graph-based methods, GEM (Fang et al., 2022), MolCLR (Wang et al., 2022b), and Graph MVP-C (Liu et al., 2021) employ approaches such as graph isomorphism networks - GIN (Xu et al., 2019) or contrastive self-supervised learning (Oord et al., 2018). KPGT (Li et al., 2022c) benefit from knowledge guided pre-training which leverages molecular descriptors and fingerprints. \\ 

There are a few general issues with graph-based models that need to be addressed. Basic GNN methods have limited expressive capabilities and do not perform well in graph isomorphism tests, e.g., the 1-Weisfeiler-Lehman (WL) test (Jin et al., 2017). Higher-order GNNs (Morris et al., 2019) were introduced to overcome such difficulties. However, increased expressiveness comes at the expense of increased computational requirements. Even though processing molecules in a three-dimensional plane improves performance, it further elevates computational complexity and expense. In addition, large-scale, clean data may not be readily available for 3D modeling. Finally, in some instances, it is reported that these models do not generalize to large molecules (Ross et al., 2022).

\section{Methodology}

\vspace{-2pt}

In SELFormer, we first trained a self-supervised transformer encoder-based chemical language model to learn high-dimensional embeddings of molecules in our dataset. After that, we fine-tuned the pre-trained model in the context of supervised learning for numerous molecular property prediction tasks by adding a classification or regression head on top of the pre-trained model architecture. Below, we explain each of these steps in detail. The overview of the SELFormer model for pre-training and fine-tuning is shown in Figure 1A and 1b, respectively. \\

SELFormer is implemented using the Python programming language and the Hugging Face Transformers library (Wolf et al., 2019), which provides a simple and consistent API for building, training, and deploying state-of-the-art machine learning models. Molecules are shared in their SMILES notations in most of the chemistry resources. Therefore, we converted SMILES notations to SELFIES representations using the SELFIES API (Krenn et al., 2020). To split the fine-tuning datasets into training, and test sets, we utilized the scaffold splitter from the Chemprop library (Yang et al., 2019).

\subsection{Pre-training}

To pre-train SELFormer (Figure 1A), we employed the ChEMBL dataset (Gaulton et al., 2017) (version 30) containing SMILES representations of 2,084,725 drug-like bioactive compounds (small molecules). Prior to pre-training, we conducted the SMILES to SELFIES conversion, which resulted in 2,084,472 unique molecules in their SELFIES notation. To speed up the conversion process, we parallelized the operation using the Pandaral.lel (https://nalepae.github.io/pandarallel/) Python library. Byte-level byte-pair encoding (BPE) (Radford et al., 2019) is employed to tokenize the SELFIES notations of the ChEMBL dataset. While defining special tokens, our tokenizer adopts the same approach as RoBERTa (Liu et al., 2019), enclosing the tokens in angle brackets and using lowercase letters (e.g., <mask>, <unk>, etc.). SELFormer is built on the RoBERTa transformer architecture (Liu et al., 2019), which utilizes the same architecture as BERT (Devlin et al., 2019), but with certain modifications that have been found to improve model performance or provide other benefits. One such modification is the use of byte-level Byte-Pair Encoding (BPE) (Radford et al., 2019) for tokenization instead of character-level BPE. Another one is that, RoBERTa is pre-trained exclusively on the masked language modeling (MLM) objective while disregarding the next sentence prediction (NSP) task. \\

To optimize our pre-trained models, we conducted a hyperparameter search using a subset of our original training dataset, composed of 100,000 molecules (with a randomized 80-20 train-validation split). Due to computational resource limitations, we followed a sequential approach, resulting in the training of nearly 100 models as opposed to the thousands that would have been trained in a full grid search. In the first hyperparameter search run, we evaluated the impact of the number of attention heads (4, 8, 12), number of hidden layers (4, 8, 12), and learning rate (1e-1, 1e-3, 1e-5, 5e-5) while keeping batch size and number of epochs constant. We selected the top-performing models based on validation loss values and carried them forward to the second hyperparameter run, in which we varied the batch size (8, 16, 32) and the number of epochs (5, 10, 25) on the best-performing models from the first run. From the models that performed best, we selected two models to proceed with full-scale pre-training and subsequent fine-tuning. We conducted our experiments on 2 NVIDIA A5000 GPUs. The total amount of pre-training time for the hyper-parameter optimization tests was around 11 days.

\begin{figure*}[h]
    \centering
    \includegraphics[width=0.9\textwidth]{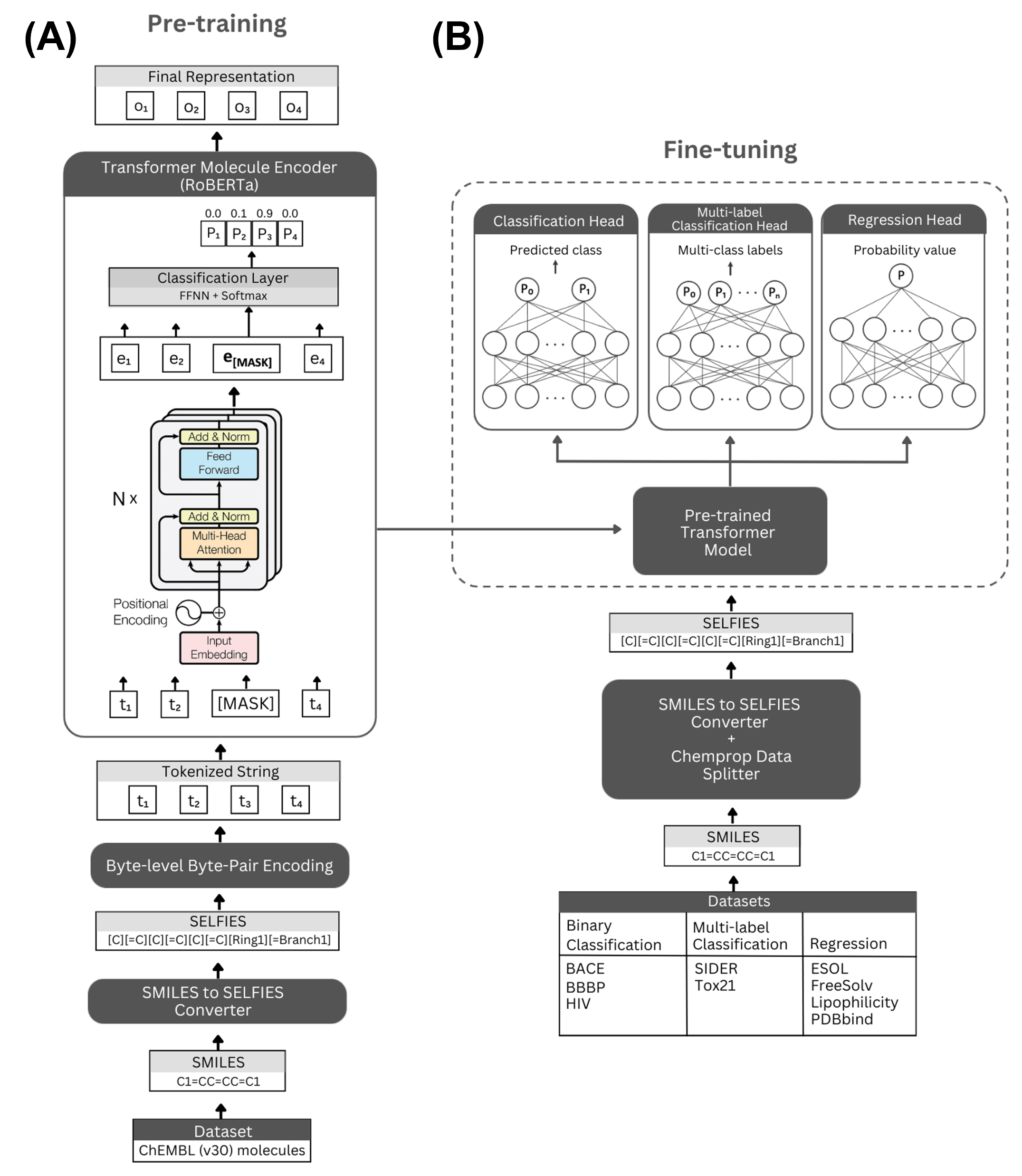}
    \caption{The schematic representation of the SELFormer architecture and the experiments conducted. \textbf{(A)} the self-supervised pre-training utilizes the transformer encoder module via masked language modeling for learning concise and informative representations of small molecules encoded by their SELFIES notation. \textbf{(B)} the pre-trained model has been fine-tuned independently on numerous molecular property-based classification and regression tasks.}
    \label{fig:img1}
\end{figure*}

\subsection{Fine-tuning}

We fine-tuned our models (Figure 1B) on several molecular property prediction tasks using the MoleculeNet (Wu et al., 2018) benchmarking datasets of varying sizes including the blood–brain barrier penetration (BBBP), the Side Effect Resource (SIDER), The “Toxicology in the 21st Century” (Tox21), the ability to inhibit HIV replication (HIV) and binding properties against the human beta-secretase 1 protein (BACE) datasets for classification tasks, and The Free Solvation Database (FreeSolv), aqueous solubility (ESOL), Lipophilicity and the binding affinity prediction (PDBbind) datasets for regression tasks. The details of the selected tasks/datasets are given in Table 1. \\ 

\begin{table}[h]
\scriptsize
\caption{Information about molecular property prediction tasks that were used in model fine-tuning.}
\centering
\vspace{0.1cm}
{\begin{tabular}{@{}lllll@{} }
\hline
\addlinespace[0.1cm]
\textbf{Dataset} & \textbf{Type of the task} & \textbf{Content} & \textbf{\# of compound}s & \textbf{Metric} \\
\addlinespace[0.1cm]
\hline
\addlinespace[0.1cm]
BBBP  & Binary classification &Binary labels on blood–brain barrier permeability & 2,039 & ROC-AUC \\
\addlinespace[0.01cm]
HIV  & Binary classification &Binary labels on the ability to inhibit HIV replication & 41,127 & ROC-AUC   \\
\addlinespace[0.01cm]
BACE & Binary classification & Binary labels on human -secretase 1 (BACE1) binding properties & 1,513 & ROC-AUC  \\
\addlinespace[0.01cm]
SIDER  & Multi-label classification & Classification of drug side-effects into 27 system organ classes & 1,427 & ROC-AUC \\
\addlinespace[0.01cm]
Tox21 & Multi-label classification & Qualitative toxicity measurements on 12 targets & 7,831 & ROC-AUC \\
\addlinespace[0.01cm]
ESOL & Regression & Aqueous solubility of common small molecules & 1,128 & RMSE  \\
\addlinespace[0.01cm]
FreeSolv & Regression & Hydration free energy of small molecules in water & 642 & RMSE  \\
\addlinespace[0.01cm]
Lipophilicity & Regression & Experimental octanol/water distribution coefficient of compounds & 4,200 & RMSE  \\
\addlinespace[0.01cm]
PDBbind & Regression & Binding affinity data for bio-molecular complexes & 11,908 & RMSE  \\
\addlinespace[0.1cm]
\hline
\end{tabular}}{}

\end{table}

To be able to fine-tune the SELFormer model for the selected downstream task, we incorporated a classification/regression head consisting of two linear layers on top of the pre-trained model (after the output of the encoder). This module takes the embedding produced by the pre-trained (encoder) model as input, and maps it to the corresponding output, a probability distribution over the target classes for the classification tasks, or a continuous value for the regression tasks.

\subsubsection{Classification tasks}

The BACE, BBBP, and HIV datasets were used for the binary classification task, while the SIDER and Tox21 datasets were employed for the multi-label classification task. The BACE dataset contains experimental quantitative (i.e., IC50) and qualitative binding properties of a set of inhibitors of human beta-secretase 1 (BACE1) protein (Subramanian et al., 2016). The dataset comprises 2-D structures and binary labels for 1,513 compounds. The BBBP dataset provides information on the blood-brain barrier permeability properties of over 2,039 compounds (Martins et al., 2012), a crucial topic for developing drugs that target the central nervous system. The HIV dataset is part of the Drug Therapeutics Program AIDS Antiviral Screen (AIDS antiviral screen data – NCI DTP data – NCI Wiki) and includes over 41,127 compounds categorized as confirmed inactive (CI), confirmed active (CA), or confirmed moderately active (CM) based on their ability to inhibit HIV replication. MoleculeNet has further modified this dataset into binary classification as inactive (CI) and active (CA and CM) compounds. Finally, the SIDER dataset is a collection of marketed drugs and their associated adverse drug reactions (Kuhn et al., 2016). MoleculeNet has classified the side effects of 1,427 approved drugs into 27 categories based on the Medical Dictionary for Regulatory Activities classification system (MedDRA). Additionally, we included the Tox21 dataset, which contains qualitative toxicity measurements for 7,831 compounds and can be used for multi-label classification on 12 different targets. To train the models on classification tasks, we utilized cross-entropy (Eq. 1) as our loss function. To assess the performance of the models, we calculated the area under receiver operating characteristic curve (ROC), and the area under the precision-recall curve (PRC). On one axis, ROC and PRC curves plot the same metric, True Positive Rate/Recall (Eq. 2), while differing in the metric plotted on the other axis, with ROC using false positive rate (Eq. 3) and PRC using precision (Eq. 4). 

\begin{equation}
L(y,\hat{y}) = -\frac{1}{N}\sum_{i}^N\sum_{c=1}^My_{ic}\log(\hat{y}_{ic})
\end{equation}

Where, \textit{N} is the number of observations, \textit{M} is the number of classes, $y_{ic}$ is the binary value indicating whether the class label \textit{c} is the correct classification for the observation \textit{i} and, $\hat{y}_{ic}$ is the predicted (softmax) probability that the observation \textit{i} is of class \textit{c}.

\begin{equation}
TPR = \frac{TP}{TP+FN}
\end{equation}

\begin{equation}
FPR = \frac{FP}{FP+TN}
\end{equation}

\begin{equation}
Precision = \frac{TP}{TP+FP}
\end{equation}

where \textit{TP}, \textit{TN}, \textit{FP} and \textit{FN} is the number of true positive, true negative, false positive and false negative predictions, respectively.

\subsubsection{Regression tasks}

ESOL, FreeSolv, Lipophilicity, and PDBbind datasets were utilized for regression-based evaluation of SELFormer. The ESOL dataset contains aqueous solubility data for 1,128 compounds along with their chemical structures (Delaney, 2004). FreeSolv, provided by the Free Solvation Database (Mobley et al., 2014) includes experimental and calculated hydration-free energy values of small molecules in water. These values are obtained from molecular dynamics simulation-based alchemical free energy calculations. The Lipophilicity dataset, curated from the ChEMBL database, provides experimental results of octanol/water distribution coefficients for 4,200 compounds, which is a critical property for drug molecules, associated to both solubility and membrane permeability. The PDBbind dataset, from the PDBbind database (Wang et al., 2004), contains experimentally measured binding affinity values for biomolecular complexes, together with 3-D Cartesian coordinates of both ligands and their target proteins derived from experimental measurements. MoleculeNet utilizes the "refined" and "core" subsets of the PDBbind database after careful processing to eliminate data artifacts. Mean squared error loss (Eq. 5) was employed as the loss function for fine-tuning the models on regression tasks. To evaluate the performance of the models, we calculated Root Mean Square Error (RMSE) (Eq. 5) score.

\begin{equation}
 RMSE = \sqrt{MSE} = \sqrt{\frac{1}{N} \sum_{i=1}^{N}(\hat{y}_i-y_i)^2}
\end{equation}

where; \textit{N} is the number of observations, $y_i$ is the true value of observation \textit{i}, and $\hat{y}_i$ is the predicted value of observation \textit{i}. \\ 

For each dataset, we prepared both random and scaffold split datasets. For both, we assumed 80\%, 10\%, and 10\% train, validation, and test split, respectively. We utilized the scaffold splitter from the Chemprop library (Yang et al., 2019), which divides molecules based on their molecular scaffold, by ensuring that the molecules containing the same scaffold do not appear in more than one split. By partitioning structurally distinct molecules into separate subsets, scaffold splitting provides a more challenging and robust evaluation of a model's learning ability than random splitting (Wu et al., 2018).

\section{Results \& Discussion}

\subsection{Pre-training and Optimization}

To optimize the pre-training procedure we gradually explored hyperparameters by focusing on different combinations, using a random subset of 100,000 molecules from the ChEMBL dataset, as detailed in section 3.1. First, we explored the impact of attention heads, hidden layers, and learning rate via random search, which led to the selection of the top performing hyperparameter values based on validation losses. Second, we examined the effect of batch size and the number of epochs on our top performing models, and found that larger batch sizes and lower number of epochs led to poor performance. We selected two best-performing models with optimal values in terms of the number of attention heads, hidden layers, batch size, and learning rate (Table 2). We named the model with fewer trainable parameters SELFormer-Lite, and the larger one, just SELFormer (the default model). Finally, we trained these two models with the full ChEMBL pre-training dataset and calculated their validation loss values as 0.328 and 0.306 for SELFormer-Lite and SELFormer, respectively. The pre-trained models are publicly accessible at https://github.com/HUBioDataLab/SELFormer for further utilization.

\begin{table}[h]
\caption{Hyperparameter configurations of the chosen SELFormer models.}
\centering
\vspace{0.2cm}
{\begin{tabular}{@{}lll@{} }
\hline
\addlinespace[0.1cm]
  & \textbf{SELFormer-Lite} & \textbf{SELFormer} \\
\addlinespace[0.1cm]
\hline
\addlinespace[0.01cm]
Batch Size  & 16 & 16\\
\addlinespace[0.01cm]
Attention Heads  & 12 & 4  \\
\addlinespace[0.01cm]
Hidden Layers & 8 & 12 \\
\addlinespace[0.01cm]
Learning Rate  & 5e-5 & 5e-5 \\
\addlinespace[0.01cm]
Weight Decay & 0.01 & 0.01 \\
\addlinespace[0.01cm]
Epoch & 100 & 100 \\
\hline
\addlinespace[0.1cm]
\# of parameters & 58,307,360 &  86,658,848 \\ 
\addlinespace[0.1cm]
\hline
\end{tabular}}{}

\end{table}

\subsection{Ablation Study}

We conducted an ablation study to investigate the impact of model fine-tuning on downstream prediction performance by comparing the predictive capabilities of two pre-trained models, SELFormer and SELFormer-Lite. We utilized randomly split versions of both the classification- and regression-based molecular property prediction datasets to measure model performances. The training time of the fine-tuned models was restricted by limiting the number of training epochs to 25 and 50, over a fixed set of hyperparameters (i.e., learning rate= 5e-5, train batch size=16, weight decay=0), to yield a basic and fair comparison between different models. In order to calculate the prediction performance of the pre-trained models without fine-tuning, we applied a one-pass learning (i.e., a single epoch training). \\ 

The results are presented in Tables 3 and 4, for classification and regression tasks, respectively. Our findings revealed that SELFormer consistently outperformed SELFormer-Lite, comparing both their pre-trained and fine-tuned versions against each other. Fine-tuning provided a significant improvement for most of the datasets. Interestingly, we observed that the base (pre-trained) models performed better than their fine-tuned counterparts on the SIDER task, which may be due to the hyperparameters not being optimal for this task. Especially in some of the tasks (e.g., BACE and Lipophilicity) 25-epoch training provided a slightly better performance compared to 50 epochs. Based on these results, we proceeded with fine-tuning experiments on a wider hyperparameter search space using SELFormer, as it demonstrated a clear superiority over SELFormer-Lite in terms of nearly all molecular property prediction tasks.

\begin{table}[h]
\footnotesize
\caption{The ablation study results. Performance of pre-trained and fine-tuned (for 25 and 50 epochs) SELFormer models on classification-based molecular property prediction tasks, using the random split datasets. Results are presented in terms of the area under receiver operating characteristic curve (ROC) and the area under the precision-recall curve (PRC) metrics.}
\centering
\vspace{0.5cm}
{\begin{tabular}{@{}lllllllllll@{} }
\hline
\addlinespace[0.1cm]
\multicolumn{1}{c}{}& \multicolumn{2}{c}{\textbf{BACE}} & \multicolumn{2}{c}{\textbf{BBBP}} & \multicolumn{2}{c}{\textbf{HIV}} & \multicolumn{2}{c}{\textbf{Tox21}} & \multicolumn{2}{c}{\textbf{SIDER}} \\
\addlinespace[0.1cm]
\hline
\addlinespace[0.1cm]
 & ROC & PRC & ROC & PRC & ROC & PRC & ROC & PRC & ROC & PRC \\
 \addlinespace[0.1cm]
\hline
 \addlinespace[0.1cm]
SELFormer-Lite-pretrain & 0.682 &
0.744 &
0.686 &
0.914 &
0.575 &
0.401 &
0.500 &
\textbf{0.533} &
0.718 &
0.831 \\
\addlinespace[0.01cm]
SELFormer-Lite-finetune-25  & 0.787 &
0.824 &
0.826 &
0.949 &
0.705 &
0.463 &
0.645 &
0.409 &
0.708 &
0.816 \\
\addlinespace[0.01cm]
SELFormer-Lite-finetune-50 & 0.773 &
0.811 &
0.829 &
0.950 &
0.703 &
0.476 &
0.650 &
0.407 &
0.717 &
0.819 \\
 \addlinespace[0.1cm]
\hline
 \addlinespace[0.1cm]
SELFormer-pretrain  & 0.699 &
0.750 &
0.727 &
0.925 &
0.600 &
0.419 &
0.523 &
0.436 &
\textbf{0.723} &
\textbf{0.831} \\
\addlinespace[0.01cm]
SELFormer-finetune-25 & \textbf{0.796} &
\textbf{0.830} &
0.857 &
0.958 &
0.703 &
0.472 &
0.641 &
0.423 &
0.699 &
0.806 \\
\addlinespace[0.01cm]
SELFormer-finetune-50 & 0.774 &
0.813 &
\textbf{0.863} &
\textbf{0.960} &
\textbf{0.714} &
\textbf{0.503} &
\textbf{0.678}&
0.489 &
0.711 &
0.814 \\
 \addlinespace[0.1cm]
\hline
\end{tabular}}{}

\end{table}

\begin{table}[h]
\caption{The ablation study results. Performance of pre-trained and fine-tuned (for 25 and 50 epochs) SELFormer models on regression-based molecular property prediction tasks, using the random split datasets. Results are presented in terms of the Root Mean Square Error (RMSE) metric.}
\centering
\vspace{0.5cm}
{\begin{tabular}{@{}lllll@{} }
\hline
\addlinespace[0.1cm]
& \textbf{ESOL} & \textbf{FreeSolv} & \textbf{Lipophilicity} & \textbf{pdbbind} \\
\addlinespace[0.1cm]
\hline
\addlinespace[0.1cm]
 & RMSE & RMSE & RMSE & RMSE \\
 \addlinespace[0.1cm]
\hline
\addlinespace[0.1cm]
SELFormer-Lite-pretrain & 1.691 &	3.124 &	1.086 &	1.515 \\
\addlinespace[0.01cm]
SELFormer-Lite-finetune-25  & 0.429 &	1.142 &	0.725 &	\textbf{1.396} \\
\addlinespace[0.01cm]
SELFormer-Lite-finetune-50 & 0.380 &	1.012 &	0.698 &	1.450 \\
\addlinespace[0.1cm]
\hline
\addlinespace[0.1cm]
SELFormer-pretrain  & 1.357 &	3.192 &	1.021 &	1.482 \\
\addlinespace[0.01cm]
SELFormer-finetune-25 & 0.407 &	1.054 &	\textbf{0.674} &	1.473 \\
\addlinespace[0.01cm]
SELFormer-finetune-50 & \textbf{0.386} &	\textbf{1.005} &	0.711 &	1.437 \\
\addlinespace[0.1cm]
\hline
\end{tabular}}{}

\end{table}

\subsection{Fine-tuning on Molecular Property Prediction}

\subsubsection{Classification Tasks}

With the aim of assessing SELFormer's generalizability and robustness, we calculated its performance on classification-based molecular property prediction tasks using both scaffold and random split datasets. Then, we compared it with the existing chemical language models and graph-based molecular representation learning methods. Our evaluation results on the scaffold and random split datasets are presented in tables 5 and 6, respectively.   \\

\begin{table}[h]
\caption{Classification-based molecular property prediction task (i.e., BACE, BBBP, HIV, Tox21 and SIDER) performance results of SELFormer and competing methods, on scaffold split datasets. Scores are given in terms of the area under receiver operating characteristic curve (ROC) metric (higher is better). Best scores for each task are shown in bold (multiple methods are marked if the difference in-between is less than 1\%). Scores were not available for the method-task combinations that are denoted with “-”. Results for D-MPNN are taken from Fang et al. (2022), and results for Hu et al., MGCN, GCN, GIN, and SchNet are taken from Wang et al. (2022b). The HIV result provided for ChemBERTa-2 was adopted from Chithrananda et al. (2020), which is the previous study of the same authors (i.e., ChemBERTa), as there were no results provided in the ChemBERTa-2 article.}
\centering
\vspace{0.5cm}
{\begin{tabular}{@{}llllll@{} }
\hline
\addlinespace[0.1cm]
& \textbf{BACE} & \textbf{BBBP} & \textbf{HIV} & \textbf{Tox21} & \textbf{SIDER} \\
\addlinespace[0.1cm]
\hline
\addlinespace[0.1cm]
 & ROC & ROC & ROC & ROC & ROC  \\
 \addlinespace[0.1cm]
\hline
 \addlinespace[0.1cm]
SELFormer & 0.832 &
\textbf{0.902} &
0.681 &
0.653 &
\textbf{0.745} \\
\addlinespace[0.1cm]
\hline
\addlinespace[0.1cm]
D-MPNN  & 0.809 & 0.710 & 0.771 & 0.759 & 0.570
\\
\addlinespace[0.01cm]
MolBERT & 0.866 & 0.762 & 0.783 & - & - 
\\
\addlinespace[0.01cm]
ChemBERTa-2  & 0.799 & 0.728 & 0.622 & - & -
\\
\addlinespace[0.01cm]
Hu et al. & 0.859 & 0.708 & 0.802 & 0.787 & 0.652
\\
\addlinespace[0.01cm]
MolCLR &  \textbf{0.890} & 0.736 & \textbf{0.806} & 0.787 & 0.652
\\
\addlinespace[0.01cm]
GraphMVP-C & 0.812 & 0.724 & 0.770 & 0.744 & 0.639
\\
\addlinespace[0.01cm]
GEM & 0.856 & 0.724 & \textbf{0.806} & 0.781 & 0.672
\\
\addlinespace[0.01cm]
MGCN & 0.734 & 0.850 & 0.738 & 0.707 & 0.552
\\
\addlinespace[0.01cm]
GCN & 0.716 & 0.718 & 0.740 & 0.709 & 0.536 
\\
\addlinespace[0.01cm]
GIN & 0.701 & 0.658 & 0.753 & 0.740 & 0.573 
\\
\addlinespace[0.01cm]
SchNet & 0.766 & 0.848 & 0.702 & 0.772 & 0.539
\\
\addlinespace[0.01cm]
KPGT & 0.855 & \textbf{0.908} & - & \textbf{0.848} & 0.649
\\
\addlinespace[0.1cm]
\hline
\end{tabular}}{}

\end{table}

SELFormer has demonstrated high performance in two of the selected classification tasks. On SIDER, which is a multi-label classification task concerning drug side effects, SELFormer outperformed all existing approaches, achieving a 10\% increase in the overall ROC values compared to the best competitor, MolCLR. In our ablation study (section 4.2), we made an interesting observation where the pre-trained model outperformed its fine-tuned counterpart on the SIDER task (i.e., the ROC values of 0.723 and 0.711, respectively). Yet, after fine-tuning the model with more optimal hyperparameter values (i.e., learning rate= 5e-5, train batch size=8, number of epochs=200, weight decay= 0.1), we observed a significant improvement in the performance of the fine-tuned model (i.e., ROC value of 0.745). On BBBP, a binary classification task for predicting brain-blood barrier permeability, SELFormer outperformed all existing approaches except KPGT. It is quite likely that the use of SELFIES representations in our models enhanced the performance in both the SIDER and BBBP tasks as a result of accurately capturing subtle structural differences between molecules (Krenn et al., 2022), which can have a significant effect on the resulting molecular property value. This is also evident from the comparison between SELFormer and ChemBERTa-2, where the latter utilizes a similar architecture to SELFormer but SMILES representations of molecules instead of SELFIES. Furthermore, SELFIES simplifies the representation of stereochemistry compared to the SMILES notation, which in turn reduces the complexity of the input data and facilitates model training. This can be particularly relevant for the prediction tasks related to drug adverse reactions, in which stereochemistry can play an important role. These findings also demonstrate the advantages of using SELFIES notation for high-throughput drug discovery and molecular design applications.\\

When compared with text-based methods on scaffold split datasets, SELFormer outperformed ChemBERTa-2 on BACE, BBBP and HIV tasks with an average of 12\% improvement over the performance of the competing method, and MolBERT on the BBBP task with an 18\% improvement (Table 5). MolBERT demonstrates a slight performance improvement over SELFormer on the BACE task, while on the HIV task, it outperforms SELFormer, possibly due to its incorporation of two supplementary pre-training objectives that consider molecular properties. Extending pre-training with domain-relevant objectives can also improve SELFormer's performance. On scaffold split-based BACE, BBBP, and SIDER tasks, SELFormer mostly dominated graph-based approaches (e.g., with an average of 20\% improvement over D-MPNN). The exceptions were for MolCLR, GEM, and KPGT especially considering the tasks of BACE, HIV and Tox21, the superior performance of which can be attributed to their pre-training on larger datasets, such as the entire ZINC (Irwin et al., 2012) and PubChem (Kim et al., 2023) databases, and higher number of trainable model parameters. Similarly, there exist a few SMILES-based chemical language models, such as MolFormer (Ross et al., 2022), that were trained on extensive datasets (i.e., 1.1 billion molecules). MolFormer achieved superior performance with a 0.937 ROC value on the BACE task, establishing a new state-of-the-art. MolFormer's performance is not fairly comparable to SELFormer, a model trained on ~2 million molecules. Nevertheless, on the SIDER task, SELFormer outperformed MolFormer (i.e., the ROC values of 0.745 and 0.690, respectively), again highlighting the benefits of employing the SELFIES notation. On the other hand, SELFormer was outperformed by all graph-based approaches on HIV and Tox21 datasets. This may be explained by the fact that these datasets are relatively larger, which prevented us from exploring a sufficiently large hyperparameter space. Increasing the size of SELFormer and exploring larger hyperparameter spaces by allocating more resources for fine-tuning can improve the model's performance.\\

\vspace{-0.5cm}
\begin{table}[h]
\caption{Classification-based molecular property prediction task (i.e., BACE, BBBP, HIV, Tox21 and SIDER) performance results of SELFormer, on random split datasets. Scores are given in terms of the area under receiver operating characteristic curve (ROC) metric (higher is better).}
\centering
\vspace{0.5cm}
{\begin{tabular}{@{}lllllllllll@{} }
\hline
\addlinespace[0.1cm]
\multicolumn{1}{c}{}& \multicolumn{2}{c}{\textbf{BACE}} & \multicolumn{2}{c}{\textbf{BBBP}} & \multicolumn{2}{c}{\textbf{HIV}} & \multicolumn{2}{c}{\textbf{Tox21}} & \multicolumn{2}{c}{\textbf{SIDER}} \\
\addlinespace[0.1cm]
\hline
\addlinespace[0.1cm]
 & ROC & PRC & ROC & PRC & ROC & PRC & ROC & PRC & ROC & PRC \\
 \addlinespace[0.1cm]
\hline
\addlinespace[0.1cm]
SELFormer & 0.836 &
0.870 &
0.863 &
0.960 &
0.714 &
0.503 &
0.641 &
0.423 &
0.745 &
0.861 \\
\addlinespace[0.1cm]
\hline
\end{tabular}}{}

\end{table}

Table 6 presents the evaluation results on the randomly split datasets. The performance of SELFormer on random split datasets is typically higher than that on scaffold split datasets due to the presence of training and test fold molecules that are similar to each other. Competing methods did not provide any results on classification tasks with random split datasets.

\subsubsection{Regression Tasks}

SELFormer was also fine-tuned on selected regression-based molecular property prediction tasks in the MoleculeNet benchmark (i.e., ESOL, FreeSolv, Lipophilicity, and Pdbbind). These tasks are evaluated based on both scaffold and random split versions of the given datasets. In Table 7, the RMSE-based performance of SELFormer is listed with one chemical language model (i.e., ChemBERTa-2) and multiple graph-based methods, for the scaffold split. SELFormer outperformed all other methods on the ESOL dataset achieving more than a 15\% improvement over the RMSE score of the best competitor (i.e., GEM). SELFIES has the ability to successfully express functional groups and aromatic systems, both of which affect the aqueous solubility of molecules (Wang et al., 2007; Bergström and Larsson, 2018). On FreeSolv and Lipophilicity tasks, SELFormer outperformed nearly the half of the graph-based models (i.e., Hu et al., GCN, GIN, MGCN, and SchNet) and dominated by the other half (i.e., D-MPNN, GEM, KPGT, GraphMVP-C, and MolCLR). ChemBERTa-2 was the only text-based (i.e., chemical language) model that could be compared with SELFormer. On the Lipophilicity prediction task, SELFormer outperformed ChemBERTa-2. \\

\begin{table}[h]
\caption{Regression-based molecular property prediction task (i.e., ESOL, FreeSolv, Lipophilicity, and pdbbind) performance results of SELFormer and competing methods, on scaffold split datasets. Scores are given in terms of the root mean squared error (RMSE) metric (lower is better). Best scores for each task are shown in bold. Scores were not available for the method-task combinations that are denoted with “-”.}
\centering
\vspace{0.5cm}
{\begin{tabular}{@{}lllll@{} }
\hline
\addlinespace[0.1cm]
& \textbf{ESOL} & \textbf{FreeSolv} & \textbf{Lipophilicity} & \textbf{pdbbind}  \\
\addlinespace[0.1cm]
\hline
\addlinespace[0.1cm]
 & RMSE & RMSE & RMSE & RMSE \\
 \addlinespace[0.1cm]
\hline
\addlinespace[0.1cm]
SELFormer & \textbf{0.682} &
2.797 &
0.735 &
1.488 \\
\addlinespace[0.1cm]
\hline
\addlinespace[0.1cm]
D-MPNN & 1.050 &
2.082 &
0.683 &
\textbf{1.397} \\
\addlinespace[0.01cm]
MolCLR & 1.110 &
2.200 &
0.650 &
- \\
\addlinespace[0.01cm]
Hu et al. & 1.220 &
2.830 &
 0.740 &
- \\
\addlinespace[0.01cm]
MGCN & 1.270 &
3.350 &
1.110 &
- \\
\addlinespace[0.01cm]
GEM & 0.798 &
\textbf{1.877} &
0.660 &
- \\
\addlinespace[0.01cm]
SchNet & 1.050 &
3.220 &
0.910 &
- \\
\addlinespace[0.01cm]
KPGT & 0.803 &
2.121 &
\textbf{0.600} &
- \\
\addlinespace[0.01cm]
GraphMVP-C & 1.029 &
- &
 0.681 &
- \\
\addlinespace[0.01cm]
GCN & 1.430 &
2.870 &
 0.850 &
- \\
\addlinespace[0.01cm]
GIN & 1.450 &
2.760 & 
0.850 &
- \\
\addlinespace[0.01cm]
ChemBERTa-2  & - &
- &
0.986 &
- \\
\addlinespace[0.1cm]
\hline
\end{tabular}}{}
\end{table}

SELFormer was compared with both language models and graph-based methods on the random split datasets (Table 8). MolBERT, Chemformer and X-MOL utilize the transformer architecture similar to SELFormer; however, they were trained using SMILES representations. X-MOL also had the advantage of being trained on a much larger dataset composed of 1.1 billion molecules curated from the ZINC database. Again, SELFormer outperformed all methods on the ESOL dataset while reporting more or less comparable results on the other tasks.

\begin{table}[h]
\caption{Regression-based molecular property prediction task (i.e., ESOL, FreeSolv, Lipophilicity, and pdbbind) performance results of SELFormer and competing methods, on random split datasets. Scores are given in terms of the root mean squared error (RMSE) metric (lower is better). Best scores for each task are shown in bold. Scores were not available for the method-task combinations that are denoted with “-”.}
\centering
\vspace{0.5cm}
{\begin{tabular}{@{}lllll@{} }
\hline
\addlinespace[0.1cm]
& \textbf{ESOL} & \textbf{FreeSolv} & \textbf{Lipophilicity} & \textbf{pdbbind}  \\
\addlinespace[0.1cm]
\hline
\addlinespace[0.1cm]
 & RMSE & RMSE & RMSE & RMSE \\
 \addlinespace[0.1cm]
\hline
\addlinespace[0.1cm]
SELFormer & \textbf{0.386} &
1.005 &
 0.674 &
1.437 \\
\addlinespace[0.1cm]
\hline
\addlinespace[0.1cm]
D-MPNN & 0.555 &
1.075 &
\textbf{ 0.555 } &
\textbf{1.391} \\
\addlinespace[0.01cm]
MolBERT & 0.531 &
\textbf{0.941 }&
 0.561 &
- \\
\addlinespace[0.01cm]
Chemformer &  0.633 &
1.230 &
 0.598 &
- \\
\addlinespace[0.01cm]
X-MOL & 0.578 &
1.108 &
 0.596 &
- \\
\addlinespace[0.1cm]
\hline
\end{tabular}}{}
\end{table}

\subsection{Use-case Analysis}

We evaluated the biochemical significance of our findings by analyzing molecular property predictions of selected molecules from downstream analysis datasets, including BBBP, SIDER, and ESOL, in which SELFormer demonstrated a high performance. As use-cases for each task, we chose extensively studied molecules with a high number of publications without considering the SELFormer prediction outcome (Figure 2). \\

BBBP is a binary classification-based molecular property prediction task comprising the ability of compounds to permeate the blood-brain barrier (BBB). The BBB serves as a selective barrier between the blood circulation and the extracellular fluid of the brain, posing a formidable challenge to the development of drugs that target the central nervous system. We selected two compounds with known effects on the nervous system: fentanyl and bromocriptine. SELformer predicted fentanyl and bromocriptine as positive and negative, respectively, in the BBBP task. Fentanyl, a potent opioid agonist that is indicated for short-term analgesia and is known to cross the BBB (Schaefer et al., 2017). Additionally, the positive BBB permeability property of fentanyl was also predicted with high probability (0.9901) by admetSAR (https://go.drugbank.com/drugs/DB00813), an open-source database that collects and curates absorption, distribution, metabolism, excretion, and toxicity (ADMET)-associated property data from published literature (Cheng et al., 2012). On the contrary, bromocriptine, a dopaminergic receptor agonist used to treat galactorrhea due to hyperprolactinemia and early Parkinson's disease, was predicted to have negative BBB permeability by admetSAR with a probability of 0.9845 (https://go.drugbank.com/drugs/DB01200), which was also predicted as negative by SELFormer. \\ 

SIDER is a multi-label classification task of grouping drug side-effects into 27 classes based on the affected system organ. We assessed the performance of SELFormer on this task by examining verapamil, a well-studied calcium channel blocker that is prescribed for angina, arrhythmia, and hypertension (https://go.drugbank.com/drugs/DB00661). The results demonstrate that SELFormer predicted several system organ classes associated with verapamil's side-effects, including infections, infestations, respiratory and gastrointestinal disorders. The SIDER database reports that verapamil commonly causes infections, constipation, and upper respiratory tract infections, with frequencies ranging from 2.4\% to 12.1\%, which are all related to system organ classes predicted by SELFormer (http://sideeffects.embl.de/drugs/2520). Cardiac failure and pulmonary oedema are other reported frequent side-effects related to vascular, respiratory, thoracic and mediastinal disorders which were also correctly predicted by SELFormer. One of the classes predicted by SELFormer for verapamil was “metabolism and nutrition disorders”, which was evaluated to be a false-positive prediction based on the annotations of this molecule in SIDER. Calcium channel blockers have been reported to inhibit insulin release, and some studies have shown reversible hyperglycemia in patients with maturity onset diabetes mellitus treated with these drugs. Also, short-term administration of calcium channel blockers to individuals with normal glucose tolerance was reported to result in a significant glucose intolerance (Russell, 1988). These findings suggest that verapamil may have metabolic effects beyond its known side-effects, although additional studies are needed to confirm their clinical relevance and mechanisms. The ability of SELFormer to predict such potential side-effects across multiple system organ classes demonstrates the model's high generalization ability for drug side-effect prediction. \\ 

\begin{figure*}[h]
    \centering
    \includegraphics[width=0.6\textwidth]{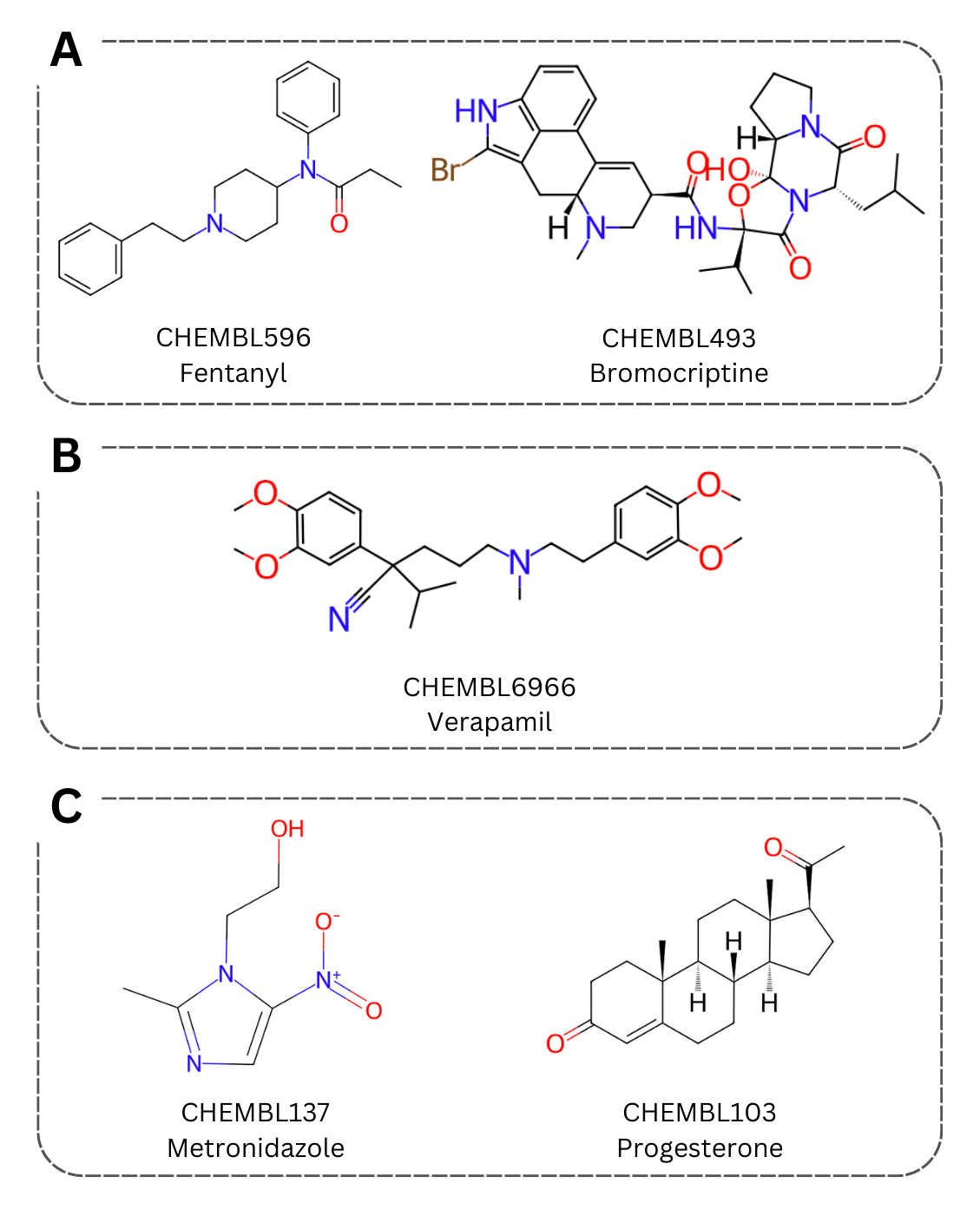}
    \caption{Selected use-case molecules from three molecular property-based datasets of; \textbf{(A)} the blood–brain barrier penetration (BBBP), \textbf{(B)} the Side Effect Resource (SIDER); and \textbf{(C)} aqueous solubility (ESOL).}
    \label{fig:img2}
\end{figure*}

The ESOL dataset is a regression task that aims to predict the solubility of compounds in water based on their chemical structures. Aqueous solubility is a key property for drug discovery and molecular design as it affects drug absorption, distribution, metabolism, and excretion in the body.  We selected two compounds, metronidazole and progesterone, to evaluate SELFormer on ESOL. Metronidazole is an antibiotic from the nitroimidazole class, commonly used to treat gastrointestinal infections, trichomoniasis, giardiasis, and amoebiasis (Dingsdag et al., 2018; Hernández Ceruelos et al., 2019). The measured log solubility of metronidazole reported in the ESOL dataset was 10.321 g/L at 25 °C (Delaney, 2004), while SELFormer predicted a value of 6.659 g/L, which represented a 35\% deviation from the measured value. Progesterone is a vital hormone for endometrial receptivity, embryo implantation, and successful pregnancy, and is also used in contraceptive preparations to prevent ovulation and fertilization and in other formulations to support pregnancy (Cooper et al., 2022). For progesterone, the measured log solubility in the ESOL dataset was 0.012 g/L, and SELFormer predicted a value of 0.028 g/L. Although SELFormer did not achieve particularly high performance on the use-cases we evaluated for the ESOL dataset, it is worth noting that aqueous solubility calculations are not always precise, leading to varying reported values across different databases. For instance, the reported aqueous solubility of metronidazole in DrugBank is 5.92 g/L (https://go.drugbank.com/drugs/DB00916), which is more consistent with SELFormer’s prediction than with the ground truth value in the ESOL dataset.

\subsection{Visualization of the embedding space}

Dimensionality reduction techniques are frequently employed to visually explore the distribution of points in a dataset, based on their features, with the aim of conducting qualitative evaluations. Here, we investigated the molecular representations produced by SELFormer via Uniform manifold approximation and projection (UMAP), which is a general purpose manifold-learning-based dimensionality reduction method (McInnes and Healy, 2008).\\ 

Due to the simplicity of visual examination and comparison between two groups, we created UMAP embeddings for the molecules in our binary classification-based molecular property prediction tasks (i.e., BACE, BBBP, and HIV). For each task, we constructed one embedding using the representations obtained from the pre-trained model, and another one using the representations obtained from the fine-tuned model. The same set of molecules were employed for the pre-trained and fine-tuned embeddings of the same task. The embeddings are given in Figure 3. \\

The molecules of the BACE task were visualized using the whole dataset since it is relatively small (i.e., composed of 1,513 data points). The two groups in Figure 3A represent the active and inactive molecules against the human beta-secretase 1 (BACE1) protein. In both pre-trained and fine-tuned model embeddings, active and inactive molecules are clearly separated from each other; however, in the fine-tuned model, two groups are slightly more linearly separable, which indicates that the fine-tuning process improved the representation power of SELFormer in the BACE task. \\ 

The BBBP dataset was embedded in Figure 3B, with the two groups representing molecules that can penetrate the blood-brain barrier and the ones that cannot. The dataset in the BBBP task is not well balanced between classes, as there is a higher number of molecules that can penetrate the brain-blood barrier. In pre-trained model embeddings, two classes are roughly separated into the two sides of the plane. Also, fine-tuning slightly improved the clustering of the non-penetrable molecules by bringing their two clusters closer to each other. \\ 

The HIV dataset contains ~41,000 molecules making it the largest classification dataset SELFormer models were fine-tuned on. This dataset is extremely unbalanced with only ~1400 positive (active) and ~39,600 negative (inactive) data points. For HIV active molecules, both pre-trained and fine-tuned model embeddings formed numerous clusters, a few of which were large and contained most of the members of the group (Figure 3C). This result shows that the models were able to capture a certain amount of structural similarity between HIV active molecules. There were small clusters of inactive molecules in the embeddings as well; however, they did not cluster well, probably due to their high number and diversity. As an overall evaluation, UMAP embeddings indicate SELFormer was able to learn the distinction between small molecules across multiple types of molecular properties, even only with pre-training. Additionally, fine-tuning causes a slight improvement. It is also important to note that, assessing the magnitude of improvement over 2-D visualizations may not be ideal, since the dimension size is not sufficient to fully express the features learned by the model.

\begin{figure*}[h]
    \centering
    \includegraphics[width=0.9\textwidth]{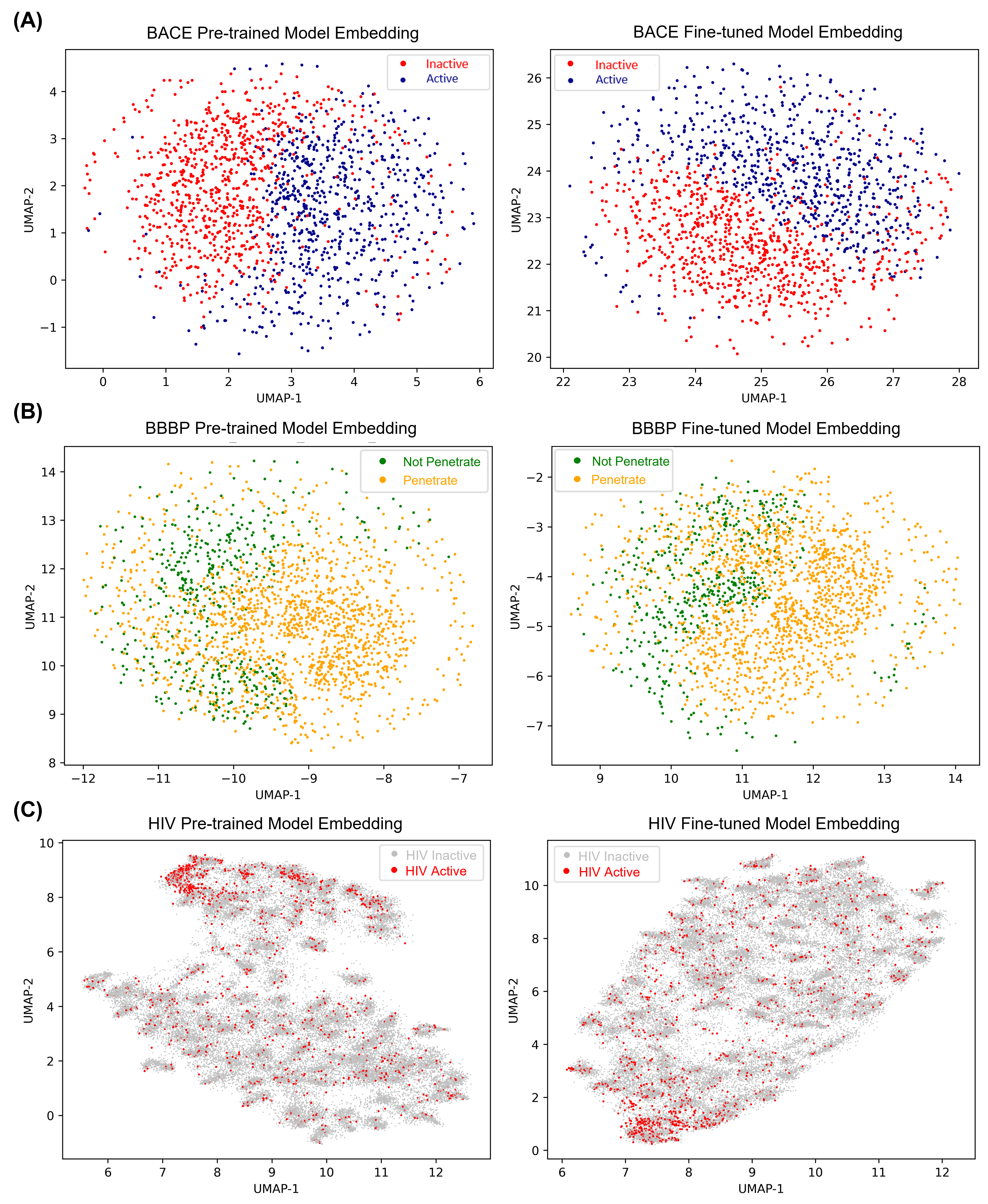}
    \caption{Visualization of SELFormer representations (via UMAP projection) considering the pre-trained (left) and task specific fine-tuned (right) SELFormer models trained on classification-based molecular property prediction benchmarks: \textbf{(A)} BACE, \textbf{(B)} the blood–brain barrier penetration (BBBP), and \textbf{(C)} HIV.}
    \label{fig:img3}
\end{figure*}

To evaluate the pre-trained SELFormer model from a drug discovery perspective, we generated a UMAP projection and 2-D visualization (Figure 4) of a large-scale molecule dataset based on compound-target interactions (i.e., bioactivity data) obtained from the ChEMBL database v29. The specific aim behind this analysis was to observe whether it is possible for SELFormer to distinguish molecules based on their functional/interaction related properties (i.e., grouping them based on their target proteins’ families), which is a highly difficult task due to the fact that molecules with dissimilar structures can interact with different members of a large protein family, and can even bind to the same region on a particular protein in some cases. The dataset we used for this experiment contains small molecule ligands (inhibitors) of proteins from five different protein families which are transferases, proteases, oxidoreductases, membrane-receptors, and ion-channels. While gathering this dataset, we only included the molecules with an experimentally measured bioactivity value of pChEMBL > 6 (i.e., IC50 < 1 uM) in single target-based binding assays. For the UMAP embedding, we randomly selected 10,000 inhibitor molecules from each protein family and eliminated molecules that are recorded to be bioactive against multiple families. Only the pre-trained SELFormer model’s representations were used in this UMAP analysis, as we did not have a model fine-tuned for this task. In Figure 4, it can be seen that molecules formed numerous small clusters, some of which are homogeneous and some are heterogeneous considering the protein families of their targets. This is most probably due to apparent structural similarities between molecules. Oxidoreductase inhibitors are clustered into several groups and located abundantly at the left hand side of the plane. Whereas, most of the protease inhibitors are located at the right-bottom side of the plot, again in several small clusters. Transferases group contains kinase inhibitors which are diverse in both structure and function, and this is reflected as scattered red dots all over the embedding plane. A similar situation applies for the membrane receptor inhibitors. \\ 

We selected sample molecules from two highly homogenous clusters (mostly composed of oxidoreductase and protease inhibitors, respectively) and one heterogeneous cluster (composed of inhibitors from all protein families) for further analysis. We calculated pairwise Bemis-Murcko scaffold-based (Bemis and Murcko, 1996) Tanimoto similarities between selected molecules that are close to each other in the UMAP embedding space. Selected oxidoreductase inhibitors had a scaffold similarity of 0.51, whereas the similarity between selected protease inhibitors was 0.42. Moreover, three molecules from the heterogeneous cluster had a mean scaffold similarity of 0.33. Their Kekule representations also indicate the structural resemblance (Figure 4). We also calculated inter-cluster similarities for the same clusters, which are far away from each other on the embedding plane. The mean pairwise scaffold similarities between the members of cluster pairs was found as; oxidoreductase vs. protease: 0.18 ± 0.04, oxidoreductase vs. heterogeneous: 0.22 ± 0.06, and protease vs. heterogeneous: 0.25 ± 0.08. These results demonstrate that SELFormer can accurately capture structural relationships, as structurally similar molecules are placed adjacently and dissimilar molecules are embedded to distant points (in different clusters) on the 2-D plane. We observed that even the heterogeneous cluster have structurally similar members and the heterogeneity mostly originates from the low correlation between structural and target interaction-related similarities. One probable explanation for the low correlation is that most of the drug candidate molecules have been screened against only one target, or a few targets from the same family. As a result, it is possible that they interact with the members of the other protein family as well (of course, this is not known presently), which would change the grouping patterns in our embedding. Overall, distinct target protein family-based clusters could not be observed mainly due to missing data and the high difficulty of this task. Nevertheless, SELFormer could capture a certain amount of functional information mostly in relation to structural similarities.

\begin{figure*}[h]
    \centering
    \includegraphics[width=0.95\textwidth]{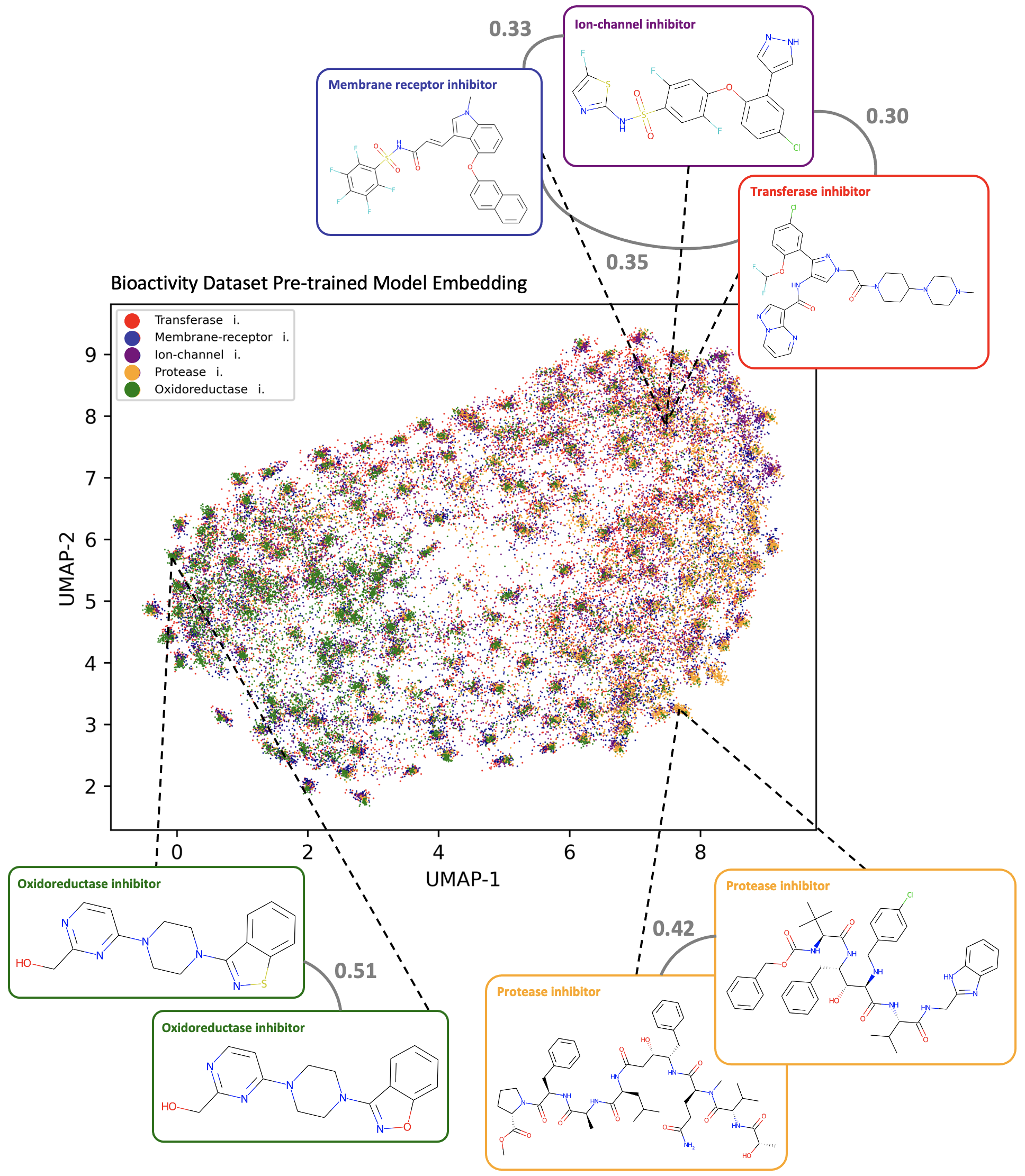}
    \caption{Visualization of SELFormer representations (via UMAP projection) of drug-like molecules from ChEMBL, grouped into high level protein families of their known targets (i.e., transferases, proteases, oxidoreductases, membrane-receptors, and ion-channels). The pairwise scaffold similarities between sample molecules (shown with dark gray colored numbers) that are embedded close to each other indicates the model's ability to construct informative molecular representations.}
    \label{fig:img4}
\end{figure*}

\section{Impact and Future Directions}

In this study, we introduced SELFormer, a new method for learning the representation of the chemical space that leverages the SELFIES notations of molecules and the transformer architecture. Our evaluation of SELFormer's performance on classification and regression-based molecular property prediction tasks revealed that it performs well and surpasses existing approaches on multiple tasks, such as predicting drug side effects, brain-blood barrier permeability, and aqueous solubility. However, larger datasets such as HIV and Tox21 presented challenges in our study mainly due to the requirement to explore large hyperparameter spaces. Comparison of SELFormer to existing language modeling-based approaches demonstrates the advantages of SELFIES notation over SMILES in molecular property prediction, including capturing diverse and complex molecular structures, simplifying stereochemistry representation, and thereby improving model accuracies. \\

The results of the ablation study suggest that careful fine-tuning of models can significantly improve performance in molecular property prediction. We further visualized the performance of the pre-trained and fine-tuned SELFormer models on binary classification-based tasks by generating UMAP embeddings. The results showed that SELFormer was able to learn the distinction between small molecules across different molecular properties, even only with pre-training. Fine-tuning the pre-trained models yielded a slight improvement in 2-D embeddings. An additional UMAP projection and visualization for a dataset based on compound-target interactions demonstrated that SELFormer was able to capture a certain amount of functional (bioactivity related) information, although distinct target protein family-based inhibitor molecule clusters could not be observed.   \\

Considering the size of our pre-training dataset and the total number of trainable parameters in the model (i.e., 2 million molecules and ~86 million parameters, respectively), it is possible to state that SELFormer is not a “large model”. Nevertheless, it performed well on the respective supervised prediction tasks. We suggest that pre-training the SELFormer model with larger molecule datasets, such as the entire ZINC database (Irwin et al., 2012) and enlarging the model by increasing the number of learnable parameters could lead to even better results, although additional experiments are needed to confirm this. \\

An interesting direction for molecular representation learning is the development of multi-modal models in which the structural data is processed together with other types of molecular information such as natural language-based text (from articles and other types of documents), and functional annotations (e.g., bio-interactions against targets, ontological associations, indications, etc.). Biomedical knowledge graphs, such as the ones provided by the CROssBAR system (Doğan et al., 2021), in which target protein interactions, signaling/metabolic pathway associations, and disease indications of drugs and millions of drug candidate compounds are represented, would be an intriguing type of data structure to leverage in this regard. \\ 

Overall, the SELFormer model exhibited high potential in molecular representation learning, which can translate well into improvements in high-throughput drug discovery and molecular design. With further exploration and optimization, SELFormer, or SELFIES-based representation models in general, could serve as a powerful and persistent alternative to SMILES-based chemical language models for encoding and analyzing complex molecular structures.

\section*{Data availability}

Code base, datasets, results and trained models of SELFormer are available at https://github.com/HUBioDataLab/SELFormer.

\section*{Author contributions}

TD conceptualized the study and designed the general methodology. AY and GD prepared the datasets, constructed the original codebase and implemented initial models. AY, EU, AU and GD trained, tuned and evaluated numerous model variants. AY constructed the finalized pre-trained SELFormer models. EU and AU performed additional data analysis and model fine-tuning on molecular property prediction, visualized the results and prepared the figures in the manuscript. AU, EU, and TD evaluated and discussed findings. AU, EU, and TD wrote the manuscript. AY, EU, AU and TD prepared the data repository. TD supervised the overall study. All authors approved the manuscript.

\section*{References}

Ahmad, W., Simon, E., Chithrananda, S., Grand, G., \& Ramsundar, B. (2022). Chemberta-2: Towards chemical foundation models. arXiv preprint arXiv:2209.01712.

\vspace{3.7pt}

AlBadani, B., Shi, R., \& Dong, J. (2022). A novel machine learning approach for sentiment analysis on twitter incorporating the universal language model fine-tuning and SVM. Applied System Innovation, 5(1), 13.

\vspace{3.7pt}

Basu, S., Ramaiah, S., \& Anbarasu, A. (2021). In-silico strategies to combat COVID-19: A comprehensive review. Biotechnology and Genetic Engineering Reviews, 37(1), 64-81.

\vspace{3.7pt}

Bengio, Y., Courville, A., \& Vincent, P. (2013). Representation learning: A review and new perspectives. IEEE transactions on pattern analysis and machine intelligence, 35(8), 1798-1828.

\vspace{3.7pt}

Bemis, G. W., \& Murcko, M. A. (1996). The properties of known drugs. 1. Molecular frameworks. Journal of medicinal chemistry, 39(15), 2887-2893.

\vspace{3.7pt}

Bergström, C. A., \& Larsson, P. (2018). Computational prediction of drug solubility in water-based systems: Qualitative and quantitative approaches used in the current drug discovery and development setting. International journal of pharmaceutics, 540(1-2), 185-193.

\vspace{3.7pt}

Brown, N., Fiscato, M., Segler, M. H., \& Vaucher, A. C. (2019). GuacaMol: benchmarking models for de novo molecular design. Journal of chemical information and modeling, 59(3), 1096-1108.

\vspace{3.7pt}

Cheng, F., Li, W., Zhou, Y., Shen, J., Wu, Z., Liu, G., Lee, P. W., \& Tang, Y. (2012). admetSAR: a comprehensive source and free tool for assessment of chemical ADMET properties. Journal of chemical information and modeling, 52(11), 3099–3105. https://doi.org/10.1021/ci300367a

\vspace{3.7pt}

Chithrananda, S., Grand, G., \& Ramsundar, B. (2020). Chemberta: Large-scale self-supervised pretraining for molecular property prediction. arXiv preprint arXiv:2010.09885.

\vspace{3.7pt}

Chuang, K. V., Gunsalus, L. M., \& Keiser, M. J. (2020). Learning molecular representations for medicinal chemistry: miniperspective. Journal of Medicinal Chemistry, 63(16), 8705-8722.

\vspace{3.7pt}

Cooper, D. B., Patel, P., \& Mahdy, H. (n.d.). Oral contraceptive pills - statpearls - NCBI bookshelf. Retrieved March 30, 2023, from https://www.ncbi.nlm.nih.gov/books/NBK430882/

\vspace{3.7pt}

Delaney, J. S. (2004). ESOL: estimating aqueous solubility directly from molecular structure. Journal of chemical information and computer sciences, 44(3), 1000-1005.

\vspace{3.7pt}

Devlin, J., Chang, M. W., Lee, K., \& Toutanova, K. (2018). Bert: Pre-training of deep bidirectional transformers for language understanding. arXiv preprint arXiv:1810.04805.

\vspace{3.7pt}

Dingsdag, S. A., \& Hunter, N. (2018). Metronidazole: an update on metabolism, structure-cytotoxicity and resistance mechanisms. The Journal of antimicrobial chemotherapy, 73(2), 265–279. https://doi.org/10.1093/jac/dkx351

\vspace{3.7pt}

Doğan, T., Atas, H., Joshi, V., Atakan, A., Rifaioglu, A. S., Nalbat, E., ... \& Atalay, V. (2021). CROssBAR: comprehensive resource of biomedical relations with knowledge graph representations. Nucleic acids research, 49(16), e96-e96.

\vspace{3.7pt}

Ericsson, L., Gouk, H., Loy, C. C., \& Hospedales, T. M. (2022). Self-supervised representation learning: Introduction, advances, and challenges. IEEE Signal Processing Magazine, 39(3), 42-62.

\vspace{3.7pt}

Fabian, B., Edlich, T., Gaspar, H., Segler, M., Meyers, J., Fiscato, M., \& Ahmed, M. (2020). Molecular representation learning with language models and domain-relevant auxiliary tasks. arXiv preprint arXiv:2011.13230.

\vspace{3.7pt}

Fang, X., Liu, L., Lei, J., He, D., Zhang, S., Zhou, J., ... \& Wang, H. (2022). Geometry-enhanced molecular representation learning for property prediction. Nature Machine Intelligence, 4(2), 127-134.

\vspace{3.7pt}

Frey, N., Soklaski, R., Axelrod, S., Samsi, S., Gomez-Bombarelli, R., Coley, C., \& Gadepally, V. (2022). Neural scaling of deep chemical models.

\vspace{3.7pt}

Gasteiger, J., Groß, J., \& Günnemann, S. (2020). Directional message passing for molecular graphs. arXiv preprint arXiv:2003.03123.

\vspace{3.7pt}

Gaulton, A., Hersey, A., Nowotka, M., Bento, A. P., Chambers, J., Mendez, D., ... \& Leach, A. R. (2017). The ChEMBL database in 2017. Nucleic acids research, 45(D1), D945-D954.

\vspace{3.7pt}

Gilmer, J., Schoenholz, S. S., Riley, P. F., Vinyals, O., \& Dahl, G. E. (2017, July). Neural message passing for quantum chemistry. In the International conference on machine learning (pp. 1263-1272). PMLR.

\vspace{3.7pt}

Handsel, J., Matthews, B., Knight, N. J., \& Coles, S. J. (2021). Translating the InChI: adapting neural machine translation to predict IUPAC names from a chemical identifier. Journal of cheminformatics, 13(1), 1-11.

\vspace{3.7pt}

Hernández Ceruelos, A., Romero-Quezada, L. C., Ruvalcaba Ledezma, J. C., \& López Contreras, L. (2019). Therapeutic uses of metronidazole and its side effects: an update. European review for medical and pharmacological sciences, 23(1), 397–401. https://doi.org/10.26355/eurrev\_201901\_16788

\vspace{3.7pt}

Hu, W., Liu, B., Gomes, J., Zitnik, M., Liang, P., Pande, V., \& Leskovec, J. (2019). Strategies for pre-training graph neural networks. arXiv preprint arXiv:1905.12265.

\vspace{3.7pt}

Irwin, J. J., Sterling, T., Mysinger, M. M., Bolstad, E. S., \& Coleman, R. G. (2012). ZINC: a free tool to discover chemistry for biology. Journal of chemical information and modeling, 52(7), 1757-1768.

\vspace{3.7pt}

Irwin, R., Dimitriadis, S., He, J., \& Bjerrum, E. J. (2022). Chemformer: a pre-trained transformer for computational chemistry. Machine Learning: Science and Technology, 3(1), 015022.

\vspace{3.7pt}

Jin, W., Coley, C., Barzilay, R., \& Jaakkola, T. (2017). Predicting organic reaction outcomes with weisfeiler-lehman network. Advances in neural information processing systems, 30.

\vspace{3.7pt}

Kalyan, K. S., Rajasekharan, A., \& Sangeetha, S. (2021). Ammus: A survey of transformer-based pretrained models in natural language processing. arXiv preprint arXiv:2108.05542.

\vspace{3.7pt}

Kim, S., Chen, J., Cheng, T., Gindulyte, A., He, J., He, S., Li, Q., Shoemaker, B. A., Thiessen, P. A., Yu, B., Zaslavsky, L., Zhang, J., \& Bolton, E. E. (2023). PubChem 2023 update. Nucleic Acids Res., 51(D1), D1373–D1380. 

\vspace{3.7pt}

Kopf, A., \& Claassen, M. (2021). Latent representation learning in biology and translational medicine. Patterns, 2(3), 100198.

\vspace{3.7pt}

Krenn, M., Ai, Q., Barthel, S., Carson, N., Frei, A., Frey, N. C., ... \& Aspuru-Guzik, A. (2022). SELFIES and the future of molecular string representations. Patterns, 3(10), 100588.

\vspace{3.7pt}

Krenn, M., Häse, F., Nigam, A., Friederich, P., \& Aspuru-Guzik, A. (2020). Self-referencing embedded strings (SELFIES): A 100\% robust molecular string representation. Machine Learning: Science and Technology, 1(4), 045024.

\vspace{3.7pt}

Kuhn, M., Letunic, I., Jensen, L. J., \& Bork, P. (2016). The SIDER database of drugs and side effects. Nucleic acids research, 44(D1), D1075–D1079. https://doi.org/10.1093/nar/gkv1075

\vspace{3.7pt}

Lewis, M., Liu, Y., Goyal, N., Ghazvininejad, M., Mohamed, A., Levy, O., ... \& Zettlemoyer, L. (2019). Bart: Denoising sequence-to-sequence pre-training for natural language generation, translation, and comprehension. arXiv preprint arXiv:1910.13461. 

\vspace{3.7pt}

Li, S., Zhou, J., Xu, T., Dou, D., \& Xiong, H. (2022). Geomgcl: Geometric graph contrastive learning for molecular property prediction. In Proceedings of the AAAI Conference on Artificial Intelligence (Vol. 36, No. 4, pp. 4541-4549).

\vspace{3.7pt}

Li, Z., Jiang, M., Wang, S., \& Zhang, S. (2022). Deep learning methods for molecular representation and property prediction. Drug Discovery Today, 103373.

\vspace{3.7pt}

Li, H., Zhao, D., \& Zeng, J. (2022). Kpgt: Knowledge-guided pre-training of graph transformer for molecular property prediction. arXiv preprint arXiv:2206.03364.

\vspace{3.7pt}

Lin, T., Wang, Y., Liu, X., \& Qiu, X. (2022). A survey of transformers. AI Open.

\vspace{3.7pt}

Liu, S., Wang, H., Liu, W., Lasenby, J., Guo, H., \& Tang, J. (2021). Pre-training molecular graph representation with 3d geometry. arXiv preprint arXiv:2110.07728.

\vspace{3.7pt}

Liu, Y., Ott, M., Goyal, N., Du, J., Joshi, M., Chen, D., ... \& Stoyanov, V. (2019). Roberta: A robustly optimized bert pretraining approach. arXiv preprint arXiv:1907.11692.

\vspace{3.7pt}

Lu, C., Liu, Q., Wang, C., Huang, Z., Lin, P., \& He, L. (2019, July). Molecular property prediction: A multilevel quantum interactions modeling perspective. In Proceedings of the AAAI Conference on Artificial Intelligence (Vol. 33, No. 01, pp. 1052-1060).

\vspace{3.7pt}

Martins, I. F., Teixeira, A. L., Pinheiro, L., \& Falcao, A. O. (2012). A Bayesian approach to in silico blood-brain barrier penetration modeling. Journal of chemical information and modeling, 52(6), 1686–1697.

\vspace{3.7pt}

McInnes, L., Healy, J., \& Melville, J. (2018). Umap: Uniform manifold approximation and projection for dimension reduction. arXiv preprint arXiv:1802.03426.

\vspace{3.7pt}

Medical Dictionary for Regulatory Activities,  MedDRA. (n.d.). Retrieved February 20, 2023, from https://www.meddra.org/ 

\vspace{3.7pt}

Mobley, D. L., \& Guthrie, J. P. (2014). FreeSolv: a database of experimental and calculated hydration free energies, with input files. Journal of computer-aided molecular design, 28(7), 711–720. https://doi.org/10.1007/s10822-014-9747-x

\vspace{3.7pt}

Morris, C., Ritzert, M., Fey, M., Hamilton, W. L., Lenssen, J. E., Rattan, G., \& Grohe, M. (2019, July). Weisfeiler and leman go neural: Higher-order graph neural networks. In Proceedings of the AAAI conference on artificial intelligence (Vol. 33, No. 01, pp. 4602-4609).

\vspace{3.7pt}

Nigam, A., Pollice, R., Krenn, M., dos Passos Gomes, G., \& Aspuru-Guzik, A. (2021). Beyond generative models: superfast traversal, optimization, novelty, exploration and discovery (STONED) algorithm for molecules using SELFIES. Chemical science, 12(20), 7079-7090.

\vspace{3.7pt}

Oord, A. V. D., Li, Y., \& Vinyals, O. (2018). Representation learning with contrastive predictive coding. arXiv preprint arXiv:1807.03748.

\vspace{3.7pt}

Radford, A., Wu, J., Child, R., Luan, D., Amodei, D., \& Sutskever, I. (2019). Language models are unsupervised multitask learners. OpenAI blog, 1(8), 9.

\vspace{3.7pt}

Ross, J., Belgodere, B., Chenthamarakshan, V., Padhi, I., Mroueh, Y., \& Das, P. (2022). Large-scale chemical language representations capture molecular structure and properties. Nature Machine Intelligence, 4(12), 1256-1264.

\vspace{3.7pt}

Russell, R. P. (1988). Side effects of calcium channel blockers. Hypertension, 11(3pt2), II42.

\vspace{3.7pt}

Schütt, K., Kindermans, P. J., Sauceda Felix, H. E., Chmiela, S., Tkatchenko, A., \& Müller, K. R. (2017). Schnet: A continuous-filter convolutional neural network for modeling quantum interactions. Advances in neural information processing systems, 30.

\vspace{3.7pt}

Schaefer, C. P., Tome, M. E., \& Davis, T. P. (2017). The opioid epidemic: a central role for the blood brain barrier in opioid analgesia and abuse. Fluids and barriers of the CNS, 14(1), 32. https://doi.org/10.1186/s12987-017-0080-3

\vspace{3.7pt}

Su, J., Lu, Y., Pan, S., Murtadha, A., Wen, B., \& Liu, Y. (2021). Roformer: Enhanced transformer with rotary position embedding. arXiv preprint arXiv:2104.09864.

\vspace{3.7pt}

Subramanian, G., Ramsundar, B., Pande, V., \& Denny, R. A. (2016). Computational Modeling of $\beta$-Secretase 1 (BACE-1) Inhibitors Using Ligand Based Approaches. Journal of chemical information and modeling, 56(10), 1936–1949. https://doi.org/10.1021/acs.jcim.6b00290

\vspace{3.7pt}

Tay, Y., Dehghani, M., Bahri, D., \& Metzler, D. (2022). Efficient transformers: A survey. ACM Computing Surveys, 55(6), 1-28.

\vspace{3.7pt}

U.S. Department of Health and Human Services. (n.d.). AIDS antiviral screen data - NCI DTP data - NCI Wiki. National Institutes of Health. Retrieved February 20, 2023, from https://wiki.nci.nih.gov/display/NCIDTPdata/AIDS+Antiviral+Screen+Data 

\vspace{3.7pt}

Unsal, S., Atas, H., Albayrak, M., Turhan, K., Acar, A. C., \& Doğan, T. (2022). Learning functional properties of proteins with language models. Nature Machine Intelligence, 4(3), 227-245.

\vspace{3.7pt}

Vamathevan, J., Clark, D., Czodrowski, P., Dunham, I., Ferran, E., Lee, G., ... \& Zhao, S. (2019). Applications of machine learning in drug discovery and development. Nature reviews Drug discovery, 18(6), 463-477.

\vspace{3.7pt}

Vaswani, A., Shazeer, N., Parmar, N., Uszkoreit, J., Jones, L., Gomez, A. N., ... \& Polosukhin, I. (2017). Attention is all you need. Advances in neural information processing systems, 30.

\vspace{3.7pt}

Wang, J., Krudy, G., Hou, T., Zhang, W., Holland, G., \& Xu, X. (2007). Development of reliable aqueous solubility models and their application in druglike analysis. Journal of chemical information and modeling, 47(4), 1395-1404.

\vspace{3.7pt}

Wang, N., Wang, Y., \& Er, M. J. (2022). Review on deep learning techniques for marine object recognition: Architectures and algorithms. Control Engineering Practice, 118, 104458.

\vspace{3.7pt}

Wang, R., Fang, X., Lu, Y., \& Wang, S. (2004). The PDBbind database: collection of binding affinities for protein-ligand complexes with known three-dimensional structures. Journal of medicinal chemistry, 47(12), 2977–2980.

\vspace{3.7pt}

Wang, Y., Wang, J., Cao, Z., \& Barati Farimani, A. (2022). Molecular contrastive learning of representations via graph neural networks. Nature Machine Intelligence, 4(3), 279-287.

\vspace{3.7pt}

Wang, S., Guo, Y., Wang, Y., Sun, H., \& Huang, J. (2019, September). SMILES-BERT: large scale unsupervised pre-training for molecular property prediction. In Proceedings of the 10th ACM international conference on bioinformatics, computational biology and health informatics (pp. 429-436).

\vspace{3.7pt}

Wigh, D. S., Goodman, J. M., \& Lapkin, A. A. (2022). A review of molecular representation in the age of machine learning. Wiley Interdisciplinary Reviews: Computational Molecular Science, 12(5), e1603.

\vspace{3.7pt}

Wolf, T., Debut, L., Sanh, V., Chaumond, J., Delangue, C., Moi, A., ... \& Rush, A. M. (2019). Huggingface's transformers: State-of-the-art natural language processing. arXiv preprint arXiv:1910.03771.

\vspace{3.7pt}

Wu, Z., Ramsundar, B., Feinberg, E. N., Gomes, J., Geniesse, C., Pappu, A. S., ... \& Pande, V. (2018). MoleculeNet: a benchmark for molecular machine learning. Chemical science, 9(2), 513-530.

\vspace{3.7pt}

Xu, K., Hu, W., Leskovec, J., \& Jegelka, S. (2018). How powerful are graph neural networks?. arXiv preprint arXiv:1810.00826.

\vspace{3.7pt}

Xue, D., Zhang, H., Xiao, D., Gong, Y., Chuai, G., Sun, Y., ... \& Liu, Q. (2020). X-MOL: large-scale pre-training for molecular understanding and diverse molecular analysis. bioRxiv, 2020-12.

\vspace{3.7pt}

Yang, K., Swanson, K., Jin, W., Coley, C., Eiden, P., Gao, H., ... \& Barzilay, R. (2019). Analyzing learned molecular representations for property prediction. Journal of chemical information and modeling, 59(8), 3370-3388.

\end{document}